\documentclass{JFM-Custom}

\usepackage{graphicx}
\usepackage{subcaption}

\newcommand{\Gtilde}{{\stackrel{\sim}{\smash{G}\rule{0pt}{1.0ex}}}}

\lefttitle{V. Iligaray, D. Aballay and F. Fuentes}

\righttitle{Global stability of 2D plane Poiseuille flow}

\title{Improved global stability bounds for two-dimensional plane Poiseuille flow}

\author{Vicente Iligaray\aff{1}, Danilo Aballay\aff{1} \and Federico Fuentes\aff{1}}

\affiliation{\aff{1} 
Institute for Mathematical and Computational Engineering (IMC), Pontificia Universidad Cat\'olica de Chile, Santiago, Chile}

\corresau{Federico Fuentes, \email{federico.fuentes@uc.cl}}

\begin{document}
\maketitle

\begin{abstract}
This work provides new lower bounds on the global (nonlinear) stability limit of pressure-driven two-dimensional plane Poiseuille flow, improving on the energy stability limit, $\Rey_E$, originally computed by Orr in 1907. 
Using a computer we carefully construct quartic Lyapunov functionals of the velocity perturbations about the laminar profile, which certify the nonlinear stability of the flow to arbitrary perturbations.
The formulation combines a decomposition of the velocity into finitely many energy eigenmodes, referred to as a `mode set', and an infinite-dimensional `tail', together with explicit bounds that recast the Lyapunov inequality conditions as semidefinite programs, whose feasibility is tested. 
Over the streamwise lengths considered, the certified stability limit exceeds the classical energy bound.
In particular, at the critical energy-stable streamwise length, where $\Rey_E\approx 87.59$, the flow is found to be globally stable up to $Re \approx 106.8$ (representing a $22\%$ improvement).
Various modestly-sized mode sets, capable of capturing sufficient features of the nonlinear dynamics of energy growth and subsequent decay, are proposed and found to be successful in producing improved bounds, with the simplest one involving only five modes.
\end{abstract}


\section{Introduction}

Pressure-driven two-dimensional (2D) plane Poiseuille flow (or channel flow) is a widely studied canonical flow in fluid mechanics \citep{orszagTransitionTurbulencePlane1980}. 
It consists of an incompressible fluid with kinematic viscosity $\nu$ governed by the Navier-Stokes equations  flowing between two parallel walls, with half-channel height $h_c$, and driven by a constant streamwise pressure gradient producing a parabolic laminar velocity profile with centreline velocity $U_c$, so that the Reynolds number is defined by $\Rey=U_c h_c/\nu$.
Despite its fundamental status, there is a lot that remains unknown about its dynamics.  

To study how this flow transitions, we analyse the stability of its steady laminar state, i.e., whether or not initial velocity perturbations about this state vanish in time. 
Its linear stability limit has been computed to be $\Rey_L\approx5772$ \citep{orszagAccurateSolutionOrr1971}, meaning that just above $\Rey_L$ there exists an initial (small) unstable perturbation which will \emph{not} decay in time.
Having said that, this number is not as informative as it seems, because transition to turbulence in this flow is subcritical, being observed well below $\Rey_L$.
Indeed, travelling waves (i.e., self-sustaining nonlaminar periodic exact solutions of the flow)  have been accurately computed as early as $\Rey\approx2939$ \citep{CASAS20122864}.
Thus, $\Rey\approx2939$ constitutes a clear upper bound to the global stability limit of the flow, $\Rey_G$, defined as the largest Reynolds number such that \textit{every} initial perturbation (no matter the magnitude) can be proved to vanish in time, resulting in the global asymptotic (nonlinear) stability of its laminar flow \citep{schmidhenningsonStabilityBook}. 
Meanwhile, the only known lower bound to $\Rey_G$ is given by its energy stability limit, i.e., the largest Reynolds number such that the kinetic energy of any perturbation monotonically decays in time. 
It was originally computed by \cite{orrStabilityInstabilitySteady1907a} to be $\Rey_E\approx88$ with a more accurate update being $\Rey_E\approx87.59$, attained by a critical periodic perturbation of nondimensional length $L_E=\tfrac{2\pi}{\alpha_E}\approx2.99$.
(The three-dimensional doubly-periodic pressure-driven version of plane Poiseuille flow shares the same linear stability limit, but has travelling waves detected at $\Rey\approx 331$ \citep{zammertStreamwiseDecayLocalized2016}, and its energy stability limit is $\Rey_E\approx 49.6$ \citep{josephStabilityPoiseuilleFlow1969b}.)

Thus, much remains unknown about the flow's global stability for $87.59\leq \Rey\leq 2939$.
The purpose of this article is to establish the global (nonlinear) stability of 2D plane Poiseuille flow beyond the energy stability limit, resulting in improved lower bounds on the global stability of the flow.
These results, settled for channels up to a certain streamwise length, represent the first global stability certificates in over a century since  \cite{orrStabilityInstabilitySteady1907a} first computed $\Rey_E$. 
Proving global stability of a fluid flow beyond $\Rey_E$ is usually a difficult endeavour that has historically been confined to seeking `quadratic' Lyapunov functionals.
Instead, we obtained these new bounds by constructing `quartic' Lyapunov functionals using a computer, following the original ideas of \cite{goulartGlobalStabilityAnalysis2012}, which were later implemented and refined by \cite{HuangFuentes} and \cite{fuentesGlobalStabilityFluid2022a}.
The computational `SOS-Lyapunov framework' searches for functionals that satisfy certain polynomial sum-of-squares (SOS) inequalities.
These are recast as a semidefinite program (SDP), whose feasibility certifies that the flow is globally stable.
(We remark that these Lyapunov functionals apply to the genuine nonlinear dynamics of the Navier-Stokes equations, whereas the results of \cite{fraternale}, upon inspection, apply only to the \emph{linearised} dynamics, because they linearised the so-called $C^1$ nonlinear compatibility conditions described by \cite{temam2006suitable} in the admissible set of velocities.)

The same general approach was used by \cite{fuentesGlobalStabilityFluid2022a} to establish the global stability of 2D plane Couette flow, but here, due to changing the flow, we had to use different numerical methods, based on finite elements, to solve the auxiliary partial differential equations (PDEs) necessary to accurately set up and apply the methodology.
Moreover, a new expression was derived to more quickly and accurately compute some of the bounds involved in the method (see Appendix \ref{app:spectral_rad}).
We note that compared to 2D plane Couette flow, where travelling waves have not been found \citep{ehrenstein2008}, 2D plane Poiseuille flow does have known upper bounds on $\Rey_G$, which makes it a richer 2D dynamical system to study.
The Lyapunov functionals have as building blocks a small set of velocity `modes', which we first need to identify. 
We make relevant observations for this specific flow, complemented with experimentation, to successfully construct several of these `mode sets' (see \S\ref{sec:results}).
Importantly, even though the computed expressions for Lyapunov functionals are often not particularly informative of the underlying physics, the selection of mode sets \emph{is} insightful and is likely to be a key ingredient for more analytical approaches to proving global stability.
Indeed, the smallest mode set we found in this work was used in the recent work of \cite{darrow2026quarticlyapunov} to analytically construct structured quartic Lyapunov functionals for this flow, though these are only valid up to certain values of $\Rey$, which are below the best bounds showcased in this work for the larger mode sets.
\section{Formulation and numerical implementation}
\label{sec:preliminaries}

For a fixed nondimensional streamwise period $L$ and Reynolds number $\Rey$, our goal is to certify global asymptotic stability of the laminar plane Poiseuille flow with respect to 2D periodic perturbations. 
We consider velocity and pressure perturbations, $\boldsymbol{u}=(u,v)$ and $p$, about the nondimensional laminar plane Poiseuille flow $\boldsymbol{U}(y)=(1-y^2,0)$ and constant pressure gradient $-\partial_x P=\tfrac{2}{Re}$ in the domain $\Omega=(0,L)\times(-1,1)$, with periodic boundary conditions in $x$ and no-slip conditions at the walls $y=\pm1$.
According to the incompressible Navier-Stokes equations, the perturbations are governed by
\begin{equation}
\partial_t \boldsymbol{u}+(\boldsymbol{u}\boldsymbol{\cdot}\nabla)\boldsymbol{u} 
+(\boldsymbol{U}\boldsymbol{\cdot}\nabla)\boldsymbol{u}
+(\boldsymbol{u}\boldsymbol{\cdot}\nabla)\boldsymbol{U}
=-\nabla p+\tfrac{1}{Re}\nabla^2\boldsymbol{u},
\qquad
\nabla\boldsymbol{\cdot}\boldsymbol{u}=0.
\label{eq:NSpert}
\end{equation}

To certify global stability, we use the SOS-Lyapunov framework developed by \cite{goulartGlobalStabilityAnalysis2012} and refined by \cite{fuentesGlobalStabilityFluid2022a}. Specifically, we seek a Lyapunov functional, $V(\boldsymbol{u})$, for the perturbation dynamics: if one can construct a functional that is positive away from the laminar state (i.e., $\boldsymbol{u}=\boldsymbol{0}$) and whose value monotonically decreases in time along every nonzero solution of \eqref{eq:NSpert}, then the perturbation must decay to zero as $t\to \infty$, implying that the laminar state is globally asymptotically stable. Therefore, the existence of such a functional for a given $\Rey$ provides a certified lower bound on $\Rey_G$ at $L$.

\subsection{SOS-Lyapunov framework}

We begin by decomposing the velocity field as,
\begin{equation}
	\boldsymbol{u}(x,y,t)=\textstyle{\sum\nolimits_{i=1}^m}a_i(t)\,\boldsymbol{u}_i(x,y)+\boldsymbol{v}(x,y,t),
	\label{eq:GalerkinSplit}
\end{equation}
where $\mathscr{U}=\{\boldsymbol{u}_1,\dots,\boldsymbol{u}_m\}$ is a finite orthonormal set of solenoidal modes, and $\boldsymbol{v}$ is an infinite-dimensional `tail' orthogonal to each $\boldsymbol{u}_i$. That is, $\langle  \boldsymbol{u}_i,\boldsymbol{u}_j\rangle = \delta_{ij}$, $\langle \boldsymbol v,\boldsymbol u_i\rangle=0$ and $\nabla \boldsymbol{\cdot}\boldsymbol{u}_i=0$ for all $i,j=1,\dots,m$, where $\delta_{ij}$ is the Kronecker delta,
\begin{equation} 
    \langle \boldsymbol{u}, \boldsymbol{v}\rangle = \textstyle{\int_\Omega}\, \boldsymbol{u}\boldsymbol{\cdot} \boldsymbol{v}\, \mathrm d\Omega \quad\text{and}\quad \|\boldsymbol u\|^2=\langle\boldsymbol u,\boldsymbol u \rangle.
\end{equation}
Thus, the time dependence of $\boldsymbol{u}$ is determined by the functions $a_i$ and $\boldsymbol{v}$. Indeed, note that the time-dependent perturbation energy, $E(t)=\tfrac{1}{2}\|\boldsymbol{u}\|^2$, can be written as
\begin{equation}
    E=\tfrac12\big(\boldsymbol{a}\boldsymbol{\cdot}\boldsymbol{a}+q^2\big)\qquad\text{with}\qquad
        q^2=\|\boldsymbol{v}\|^2\quad\text{and}\quad
        \boldsymbol{a}=(a_1,\dots,a_m)^{\mathrm T}\,.
\end{equation}
We consider quartic functionals of the form
\begin{equation}
V(\boldsymbol{a},q)=E(\boldsymbol{a},q)^2+P(\boldsymbol{a},q),
\label{eq:Vansatz}
\end{equation}
where $P$ is a cubic polynomial, with no constant or linear terms and only even powers of $q$. 
(One could consider more general ans\"atze of the form $V=E^d+P$ for $d>2$ and $P$ of at most degree $2d-1$, but the resulting computational cost is significantly  higher, so we focus on the $d=2$ case.)
We call $V$ a Lyapunov functional if
\begin{equation}
    V(\boldsymbol{a},q)\ge 0\quad\text{and}\quad
    \mathcal{L}V(\boldsymbol{a},q)\le 0
    \quad\,\,
    \text{with equality if and only if }(\boldsymbol{a},q)=(\boldsymbol{0}, 0).
    \label{eq:Lyapdef}
\end{equation}
Here $\mathcal LV(\boldsymbol a(t),q(t))=\tfrac{\mathrm dV}{\mathrm dt}(\boldsymbol a(t),q(t))$ denotes the Lie derivative of $V$ along solutions of \eqref{eq:NSpert} written in the form \eqref{eq:GalerkinSplit}. 
The existence of such a (radially unbounded) functional implies global asymptotic stability of the laminar state \citep[Prop.~3.2]{mironchenkoNoncoerciveLyapunovFunctions2019}.
Importantly, these conditions are sufficient to establish the global stability of the \emph{infinite-dimensional} system \eqref{eq:NSpert}, and not merely that of a truncated finite-dimensional approximation.
This is because the full perturbation velocity field $\boldsymbol{u}$ is properly encoded in $(\boldsymbol a,q)$, and, when taking into account the dynamics, $\mathcal LV$ includes the time-evolution of both $\boldsymbol a(t)$ and $q(t)$.

For example, consider the energy $E$, which is positive definite. 
If additionally, for any possible perturbation governed by the Navier-Stokes equations \eqref{eq:NSpert}, its energy satisfies $\tfrac{\mathrm dE}{\mathrm dt}\leq -cE$ for some $c>0$ (which occurs precisely when $\Rey\leq\Rey_E$), Gr\"onwall's inequality yields $E(t)\leq E(0)e^{-ct}$, so \emph{any} perturbation, regardless of its initial magnitude or form, eventually vanishes in time, resulting in a globally stable flow.
In these cases, the energy is a Lyapunov functional.
This can be reformulated as polynomials, since $E(\boldsymbol{a},q)=\tfrac{1}{2}(\boldsymbol{a}\,\!\boldsymbol{\cdot}\,\!\boldsymbol{a}+q^2)$ is a quadratic polynomial in $(\boldsymbol{a},q)$, which will be a Lyapunov functional at the Reynolds numbers for which $\mathcal{L}E(\boldsymbol{a},q)<0$ at every nonzero $(\boldsymbol{a},q)$.

The idea is to find more general quartic Lyapunov functionals for $\Rey>\Rey_E$.
In the ansatz \eqref{eq:Vansatz} the leading term $E^2$ guarantees radial unboundedness (meaning that $V\to\infty$ when \mbox{$\boldsymbol{a}\boldsymbol{\cdot}\boldsymbol{a}+q^2\to\infty$}) and that $\mathcal{L}V$ is of even degree, while the lower-degree polynomial $P$ provides the flexibility to enforce the inequalities in \eqref{eq:Lyapdef} with the aid of a computer (see \cite{goulartGlobalStabilityAnalysis2012} and \cite{fuentesGlobalStabilityFluid2022a} for more details).

Projecting \eqref{eq:NSpert} onto $\operatorname{span}(\mathscr U)$ and its orthogonal complement yields equations for ${\mathrm d \boldsymbol{a}}/{\mathrm d t}$ and ${\mathrm d q^2}/{\mathrm d t}$. These equations are not closed in $(\boldsymbol a,q)$, since some terms depend on the tail $\boldsymbol v$ beyond its energy $\tfrac{1}{2}q^2$. Following \cite{fuentesGlobalStabilityFluid2022a}, such terms can nevertheless be conveniently bounded when $\mathscr U$ is chosen as a finite set of `energy eigenmodes' (described in \S\ref{sec:modes}). This leads to explicit sufficient conditions for \eqref{eq:Vansatz} to define a Lyapunov functional satisfying \eqref{eq:Lyapdef}. Specifically, global stability of the flow is verified by finding a cubic polynomial $P$ with no constant or linear terms, quartic polynomials $r_i$ with no constant or linear terms, and quadratic polynomials $s_i$, all involving only even powers of $q$, such that
\begin{equation}
\begin{gathered}
E(\boldsymbol{a},q)^2 + P(\boldsymbol{a},q) - \varepsilon E(\boldsymbol{a},q) \ge 0,\\
-\big(\Gtilde(\boldsymbol{a},q)+\textstyle{\sum_{i=1}^m}\big(r_i(\boldsymbol{a},q)+C_i q^2 s_i(\boldsymbol{a},q)\big)+\varepsilon E(\boldsymbol{a},q)\big) \ge 0,\\
\frac{\partial V}{\partial q^2} \ge 0,\quad s_i(\boldsymbol{a},q)-M_i(\boldsymbol{a},q) \ge 0, \quad
s_i(\boldsymbol{a},q)+M_i(\boldsymbol{a},q) \ge 0, \quad i=1,\dots,m,\\
\begin{bmatrix} w_1 & w_2 \end{bmatrix}
\begin{bmatrix}
q^2\,\tilde{\boldsymbol{a}}^{\mathrm T}\mathsfbi{G}_i\tilde{\boldsymbol{a}}\,r_i(\boldsymbol{a},q) &
q^2\,\tilde{\boldsymbol{a}}^{\mathrm T}\mathsfbi{G}_i\tilde{\boldsymbol{a}}\,M_i(\boldsymbol{a},q) \\
q^2\,\tilde{\boldsymbol{a}}^{\mathrm T}\mathsfbi{G}_i\tilde{\boldsymbol{a}}\,M_i(\boldsymbol{a},q) &
r_i(\boldsymbol{a},q)
\end{bmatrix}
\begin{bmatrix} w_1 \\ w_2 \end{bmatrix}
\ge 0,\quad i=1,\dots,m.
\label{eq:SOScompact}
\end{gathered}
\end{equation}
for all $(\boldsymbol{a},q,w_1,w_2)\in\mathbb{R}^{m+3}$ where $\tilde{\boldsymbol{a}} = (1, a_1,\dots,a_m)^{\mathrm{T}}$ and $\varepsilon>0$. Here $\smash{\Gtilde}$ and $M_i$ are polynomials, $\mathsfbi{G}_i$ are constant tensors, and $C_i$ are constants defined in Appendix~\ref{sec:sos_system}. 

Note that all the terms in \eqref{eq:SOScompact} are polynomials. 
For a fixed choice of $\Rey$, $L$ and $\mathscr U$, to reliably solve \eqref{eq:SOScompact} with a computer, the inequalities are strengthened to SOS polynomial constraints and recast as an SDP which can be solved numerically. 
If the SDP is found to be feasible, its numerical solution includes the explicit coefficients of monomials in $P(\boldsymbol{a},q)$, with the resulting $V(\boldsymbol{a},q)$ in \eqref{eq:Vansatz} being a Lyapunov functional satisfying \eqref{eq:Lyapdef} (see \eqref{eq:example} for an explicit example).

\subsection{Energy-rate eigenvalue problem}\label{sec:modes}

The modes in $\mathscr U$ are chosen from the solutions to the 2D second-order energy-rate eigenvalue problem in $\Omega$ \citep[\S2.2]{doeringAppliedAnalysisNavierStokes1995b},
\begin{equation}
\tfrac{1}{Re} \nabla^2 \boldsymbol{u} -\tfrac12\big(\nabla \boldsymbol{U} + (\nabla\boldsymbol U)^{\mathrm T}\big)\boldsymbol{u}-\nabla \zeta = \lambda \boldsymbol u,
\qquad
\nabla\boldsymbol{\cdot}\boldsymbol{u}=0, \qquad \|\boldsymbol{u}\|^2 = 1\,.
\label{eq:energy2D}
\end{equation}
Here, $\zeta$ is an $x$-periodic Lagrange multiplier enforcing incompressibility, $\boldsymbol{u}$ satisfies periodic boundary conditions in $x$ and no-slip at $y=\pm 1$, and $\lambda=\tfrac{\mathrm dE}{\mathrm dt}\big|_{t=0}$ is the instantaneous energy growth rate of \eqref{eq:NSpert} when the `energy eigenmode' $\boldsymbol{u}$ is chosen as the initial condition. 
Since the problem is two-dimensional, we represent the velocity in terms of a streamfunction $\psi$, writing $\boldsymbol u=(\partial_y \psi,-\partial_x\psi)$, which automatically satisfies the incompressibility constraint in \eqref{eq:energy2D}, while the third equation is enforced by a posteriori rescaling of the eigenmodes. 

Letting $\psi(x,y) = \sum_{n\in\mathbb Z} \varphi_n(y) e^{\text{i}\alpha_n x}$ with $\alpha_n = \frac{2\pi n}{L}$, and taking the curl of the first equation in \eqref{eq:energy2D} yields a one-dimensional $\mathbb{C}$-valued fourth-order generalised Hermitian eigenvalue problem for each $n$. 
For $\alpha\neq 0$ (dropping the $n$ for convenience), 
\begin{equation}
\textstyle{\frac{1}{Re}}\bigl(\alpha^4\varphi-2\alpha^2\varphi''+\varphi''''\bigr)-\text{i}\alpha\bigl(2y\varphi'+\varphi\bigr)
\!=\!
\lambda\bigl(\varphi''-\alpha^2\varphi\bigr)\,\,\,\text{with}\,\,\,\varphi(\pm1)\!=\!\varphi'(\pm1)\!=\!0.
\label{eq:energyEVP}
\end{equation}
For each eigenvalue $\lambda$, there are two linearly independent real-valued velocity eigenmodes, chosen to be orthogonal via $\psi_{\!A}(x,y)=\mathfrak{R}\big(e^{\text{i}\theta}\varphi(y)e^{\text{i}\alpha x}\big)$ and $\psi_{\!B}(x,y)=\mathfrak{R}\big(\text{i}e^{\text{i}\theta}\varphi(y)e^{\text{i}\alpha x}\big)$ (thus, mutually streamwise shifted by $\tfrac{L}{4n}$). 
Here, $|z|$, $\mathfrak{R}(z)$ and $\bar z$ denote the modulus, real part, and complex conjugate of $z\in\mathbb{C}$ respectively.
Any fixed $\theta$, representing a streamwise shift, will work, but we select it such that $e^{\text{i}\theta}\big(\varphi(0) + \text{i}\varphi'(0)\big)=|\varphi(0)| + |\varphi'(0)|$, which ensures $\psi_{\!A}$ has a stagnation point at the origin.

For $\alpha = 0$ the analytical solution are the eigenvalues $\lambda=-\tfrac{1}{4Re}(k+1)^2\pi^2$, each having a single velocity eigenmode $\psi(y)=\mathfrak{R}\big(e^{\text{i}(\pi/2)(k+1)}e^{\text{i}(\pi/2)(k+1)y}\big)$, where $k$ is a nonnegative integer. 
Indeed, this solution comes from a simpler second-order eigenvalue problem for the eigenmodes $\boldsymbol{u}=(u_0,0)=(\partial_y\psi,0)$, directly derived from \eqref{eq:energy2D} when $\alpha=0$, namely
\begin{equation}
    \tfrac{1}{Re}u_0'' = \lambda u_0\qquad \text{with}\qquad u_0(\pm1)=0.
    \label{eq:energyEVP0}
\end{equation}

\subsection{Numerical methods and computational implementation} \label{sec:nummethod}

To implement the inequalities in \eqref{eq:SOScompact} we must first numerically solve \eqref{eq:energy2D}, select a finite subset of energy eigenmodes comprising $\mathscr{U} = \{\boldsymbol{u}_1,\ldots, \boldsymbol{u}_m\}$ (selection criteria are discussed in \S\ref{sec:results}), and then compute the relevant tensors in the polynomials $\smash{\Gtilde}$, and the terms $\mathsfbi{G}_i$ and $C_i$ for $i=1,\ldots,m$, which are specified in Appendix \ref{sec:sos_system}.
 
To numerically solve \eqref{eq:energyEVP} we use finite element methods (FEM). 
With this in mind, we multiply by a sufficiently regular test function and integrate by parts accordingly, yielding a variational formulation of the form
\begin{equation}
    \text{find}\,\,\varphi\in \mathscr{V}\,\,\text{such that}\quad  
    \mathscr{a}(\varphi, \eta) = \lambda\mathscr{b}(\varphi,\eta)\quad \forall \eta\in \mathscr{V}.
\end{equation}
Here, $\mathscr{V}$ is chosen as the Sobolev space $H_0^2(-1,1)$, which is well approximated by piecewise polynomial functions $\varphi$ with continuous derivative satisfying $\varphi(\pm1)=\varphi'(\pm1)=0$, and 
\begin{equation}
\begin{aligned}
    \mathscr{a}(\varphi, \eta) &= \textstyle{\frac{1}{Re}\int_{-1}^1}
\big(
\varphi''\overline{\eta''}
+2\alpha^2\varphi'\overline{\eta'}
+\alpha^4\varphi\overline{\eta}
\big)\,\mathrm{d}y
-\text{i}\alpha\int_{-1}^1
(2y\varphi'+\varphi)\overline{\eta}\,\mathrm{d}y,\\
\mathscr{b}(\varphi,\eta) &= -\textstyle{\int_{-1}^1}
\big(
\varphi'\overline{\eta'}
+\alpha^2\varphi\overline{\eta}
\big)\,\mathrm{d}y\,.
\end{aligned}
\label{eq:weakEVP}
\end{equation}

To discretize $\mathscr{V}$ we choose an $H^2$-conforming discretization subordinate to a uniform mesh of $\Omega_y = (-1,1)$ with element size $h$, comprised of continuously differentiable locally supported cubic Hermite finite elements.
This leads to a generalised eigenvalue problem of the form $\mathsfbi{A}\mathsf{z} = \lambda \mathsfbi{B}\mathsf{z}$ where $\mathsfbi A$ and $\mathsfbi B$ are Hermitian matrices. 
For each $\alpha_n$, we assemble these matrices and solve the problem using an in-house MATLAB \citep{matlab} implementation with mesh size $h=0.001$, and then construct the velocity eigenmodes as described previously in \S\ref{sec:modes}. 
The largest energy eigenvalue not associated with the eigenmodes in $\mathscr{U}$ is labelled $\kappa$ and recorded for future use (see \eqref{eq:polySOSformulation}). 

The relevant tensors in $\smash{\Gtilde}$ (namely, $\mathsfbi L$ and $\mathsfbi N$ in \eqref{eq:LNtensors}) involve integrals whose values are computed numerically using quadrature as they only entail piecewise polynomial finite element integrands. 
The matrices $\mathsfbi G_i$ are computed as described in Appendix~\ref{app:solenoidal_projection}, while the bounds $C_i$ are computed using a new accurate procedure detailed in Appendix~\ref{app:spectral_rad}.
Equipped with these quantities, and setting $\varepsilon=10^{-5}$, we are able to set up the SDP associated with \eqref{eq:SOScompact} in MATLAB using YALMIP \citep{Lofberg2004}.
We then determine its feasibility by solving with MOSEK v8.0.0.81 \citep{mosek}. 
Given $\mathscr{U}$, for each $L$ we recorded the largest $\Rey$, to within $0.025$, with a valid quartic Lyapunov functional. 
For other details of the numerical implementation and its verification, see Appendix~\ref{app:implementation}.
\vspace{-4mm}
\section{Results and discussion}\label{sec:results}

\subsection{Selection of mode sets}

The physical reasoning behind this whole framework relies on selecting a relatively small explicit `mode set' $\mathscr U$, which should at least capture the nonlinear dynamics of any existing energy growth in the flow and, left to its own means (i.e., ignoring the tail $\boldsymbol{v}$ in \eqref{eq:GalerkinSplit}), be able to globally stabilise such energy growth, eventually dissipating it entirely.
In other words, the dynamics of \eqref{eq:NSpert} truncated to $\mathscr U$ (with reference to \eqref{eq:GalerkinSplit}, given by $\tfrac{\mathrm d a_i}{\mathrm d t}=\sum_{j=1}^m\mathsfbi{L}_{ij}a_j+\sum_{j,k=1}^m\mathsfbi{N}_{ijk}a_j a_k$ for $\mathsfbi{L}$ and $\mathsfbi{N}$ in \eqref{eq:LNtensors}) ought to be globally stable.
That said, the residual, represented by the tail $\boldsymbol{v}$ in \eqref{eq:GalerkinSplit} and whose isolated dynamics should have monotonically decreasing energy (i.e., $\tfrac{\mathrm d}{\mathrm d t}(\tfrac{1}{2}q^2)=\tfrac{1}{Re}\langle \boldsymbol{v},\nabla^2 \boldsymbol{v}\rangle - \langle \boldsymbol{v},(\nabla \boldsymbol{U}+(\nabla\boldsymbol U)^{\mathrm T})\boldsymbol{v}\rangle<0$), is essential to yield rigorous global stability results for the complete dynamics described by \eqref{eq:NSpert}.
Thus, a careful selection of $\mathscr U$ is fundamental for the methodology to work.

In line with the reasoning above, two quick preliminary checks are essential when selecting the energy eigenmodes comprising $\mathscr U$.
First, the largest energy eigenvalue among the omitted modes, which we refer to as $\kappa$, must be negative. 
That is, \emph{all} the energy eigenmodes with instantaneous energy growth (i.e., $\lambda>0$ in \eqref{eq:energy2D}) at given $L$ and $\Rey$ should be present in $\mathscr U$.
Second, the truncated dynamics of \eqref{eq:NSpert} to $\operatorname{span}(\mathscr U)$ must be linearly stable (i.e., eigenvalues of $\mathsfbi{L}$ in \eqref{eq:LNtensors} should all have negative real part). 
Having said that, these conditions are not sufficient for \eqref{eq:SOScompact} to hold, so careful experimentation is required to settle on mode sets $\mathscr U$ with nontrivial interactions resulting in global stability.

\begin{figure}
	\centering
	\includegraphics[width=0.96\textwidth]{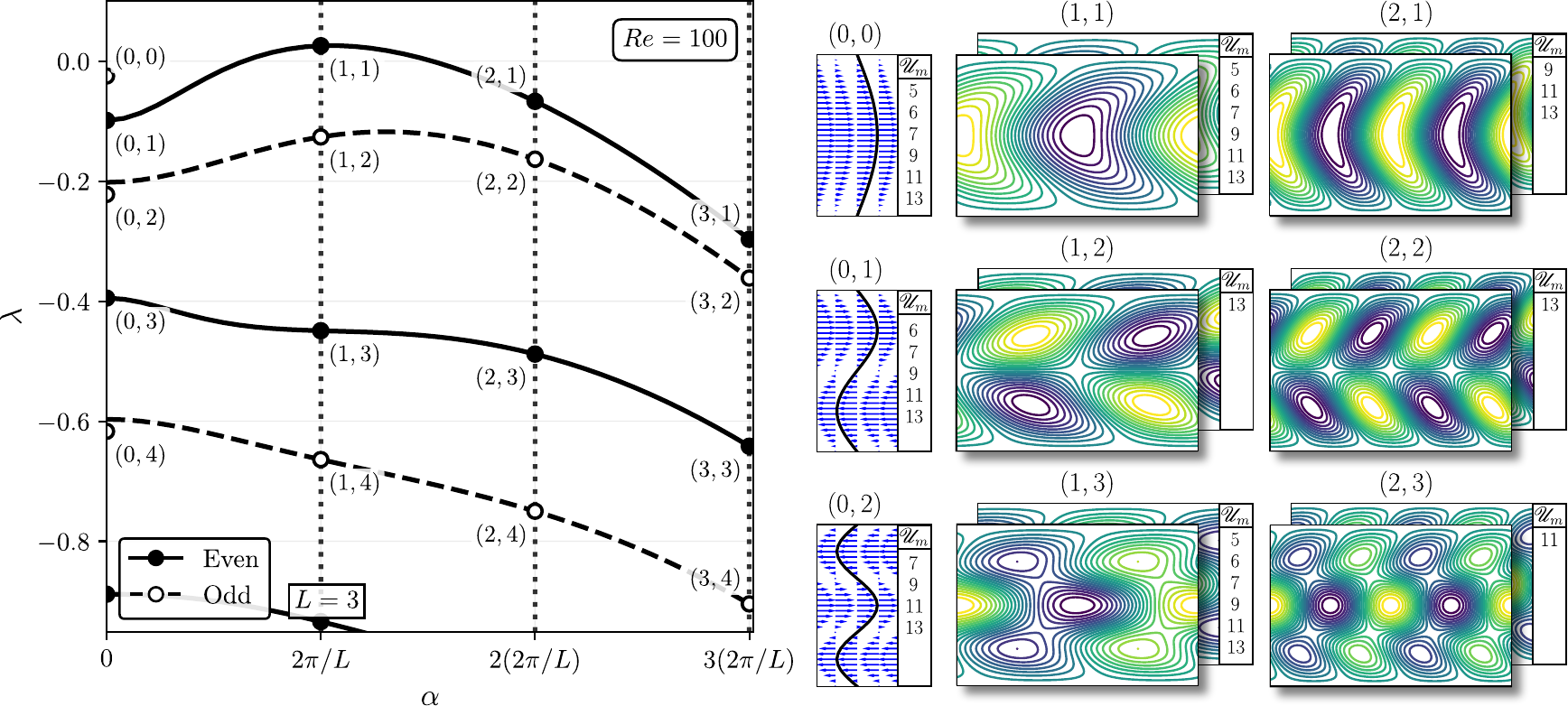}
	\caption{
		The left panel shows the energy eigenvalue branches of 2D Poiseuille flow at $Re=100$ as a function of the streamwise wavenumber $\alpha$, labelled by the parity of the corresponding streamfunction. 
		Eigenvalues consistent with $L=3$ are labelled as $(n,k)$, which, when $n\neq0$, correspond to the $k$-th largest eigenvalues at the fixed wavenumber $\alpha_n=\tfrac{2\pi}{L}n$, and when $n=0$ correspond to $\lambda=-\tfrac{1}{4Re}(k+1)^2\pi^2$ for $k\geq0$. 
		The right panels show the corresponding eigenmodes, including their multiplicity, along with a list of the modes sets $\mathscr{U}_m$ to which they belong to.
		\vspace{-4mm}
	}
	\label{fig:EnergyEVP}
\end{figure}

In this work, we selected several mode sets, labelled for simplicity as $\mathscr U_m$ with $m$ coinciding with its cardinality.
In what follows, we provide some insight into the choice of the $\mathscr U_m$.
To facilitate the discussion, it is useful to look at the energy spectrum in detail and try to understand the eigenmode interactions.
As a guide, see Figure \ref{fig:EnergyEVP}, which illustrates the energy eigenvalues and eigenmodes when $L=3$ and $\Rey=100$.
Eigenvalue branches are clearly observed, so, at the wavenumber $\alpha_n=2\pi n/L$ for $n\neq0$, we label the energy eigenvalue as $(n,k)$ if it sits in the $k$-th branch for $k=1,2,\ldots$, and recall that there are two linearly independent eigenmodes associated with it (see \S\ref{sec:modes}), clearly depicted in Figure~\ref{fig:EnergyEVP}.
Meanwhile, the $(0,k)$ modes are a special case, labelled from $k=0$, with the eigenvalues being $\lambda=-\tfrac{1}{4Re}(k+1)^2\pi^2$ and having a single streamwise-independent energy eigenmode.
Moreover, the eigenvalue branches have been classified by parity, since, along them, the streamfunctions associated with the eigenmodes are either odd or even about $y=0$ (e.g. the first branch has even streamfunctions satisfying $\psi(x,-y)=\psi(x,y)$).
This parity is caused by the symmetries of Poiseuille flow, and was first pointed out by Orr \citeyearpar[p. 76]{orrStabilityInstabilitySteady1907a}.

To hone in on a minimal mode set leading to an improvement on energy stability, let $L=3$ and consider a $\Rey>\Rey_E\approx87.6$.
As seen in Figure \ref{fig:EnergyEVP}, the $(1,1)$ mode is the only one leading to energy growth (this is true up to $\Rey\approx124$), so it \emph{must} be in $\mathscr U_m$ (and will lead to $\kappa<0$).
Now, to have linear stability within $\mathscr U_m$, there needs to be another mode that linearly interacts and stabilises the $(1,1)$ modes, so that, at the very least, the corresponding nondiagonal entry $\mathsfbi{L}_{ij}$ is nonzero (see \eqref{eq:LNtensors}).
All modes of the form $(n,k)$ for $n\neq1$ are naturally orthogonal with the $(1,1)$ modes, so they do not linearly couple, and, because of different parity, the same is true for the $(1,k)$ modes for even $k$.
Thus, the first modes that linearly couple with the $(1,1)$ modes are the $(1,3)$ modes, which are fortunately enough to linearly stabilise them.
Lastly, we add the $(0,0)$ mode as well, yielding the minimal $5$-mode set $\mathscr U_5$.
The $(0,k)$ modes, which always have decreasing energy, do not interact linearly with the other modes, but they do, however, couple nonlinearly (through the quadratic tensor $\mathsfbi N$ in \eqref{eq:LNtensors}), representing a genuinely nonlinear effect, typically stabilising, which appears to be necessary to obtain nontrivial quartic Lyapunov global stability certificates. This claim is due to methodical experimentation with both the full system and the truncated system obtained when the tail is ignored.
The discussion of the larger mode sets is continued below after looking at the initial results.

\subsection{New global stability bounds} 

\begin{figure}
	\centering
	\includegraphics[width=0.7\textwidth,trim=0 5 0 5, clip]{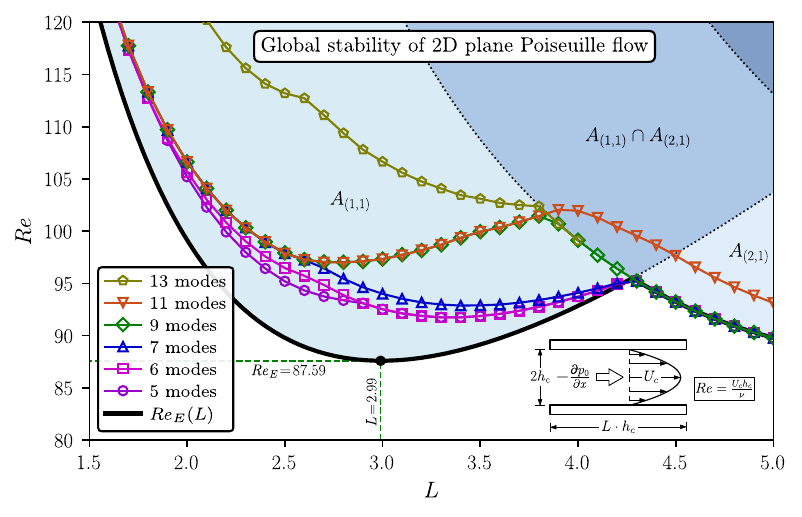}
	\caption{Global stability curves resulting from using the SOS-Lyapunov framework with the mode sets $\mathscr{U}_m$ for $m=5,6,7,9,11,13$ (see Figure \ref{fig:EnergyEVP} for details), together with the energy stability limit $\Rey_E(L)$. 
		For each $L$, plotted are the largest Reynolds numbers for which a quartic Lyapunov functional was found certifying global stability of 2D plane Poiseuille flow.
		Regions where the $(n,k)$ energy eigenvalues are positive (see Figure \ref{fig:EnergyEVP}) are shaded and labelled as $A_{(n,k)}$ accordingly.
		\vspace{-3mm}}
	\label{fig:stability}
\end{figure}

The energy stability limit as a function of $L$, denoted by $\Rey_E(L)$, which attains its minimum of $\Rey_E\approx 87.59$ at $L_E\approx 2.99$, is plotted in Figure \ref{fig:stability} along with the global stability results using $\mathscr U=\mathscr U_m$ for the different mode sets. 
Focusing on $\mathscr U_5$ first, we see that our construction yields a moderate yet visible improvement on $\Rey_E(L)$ for $1.5<L<4.3$, where only the $(1,1)$ modes produce energy growth.
Indeed, the regions where the $(n,k)$ modes lead to energy growth are very informative, so are shaded and labelled as $A_{(n,k)}$ in Figure \ref{fig:stability}. 
For $L>4.3$, the results of $\mathscr U_5$ coincide with those of $\Rey_E(L)$, which is to be expected: energy growth at these $L$ is due to the $(2,1)$ modes, so to get a nontrivial improvement, the $(2,1)$ modes (which are absent from $\mathscr U_5$) should be included in $\mathscr U$.

Adding the $(0,1)$ and $(0,2)$ modes successively to $\mathscr{U}_5$ leads to $\mathscr{U}_6$ and $\mathscr{U}_7$ respectively.
As can be seen from Figure \ref{fig:stability}, these $n=0$ modes provide further small improvements in stability, but to see more notable improvements, other $n\neq0$ modes have to be added to the mix. 
The addition of the $(2,1)$ modes in $\mathscr{U}_9$, for example, produces much better results, which, as expected, stop abruptly at the boundary of $A_{(2,1)}$, when those modes yield energy growth and require of other modes to linearly stabilise their effects.
 Inclusion of the $(2,3)$ modes in $\mathscr{U}_{11}$ does precisely that and results in global stability even in $A_{(1,1)}\cap A_{(2,1)}$, where four modes grow in energy.
 In truth, reasonable and patient manual testing of mode sets by trial and error is important: just the right combination of modes might result in a significant improvement.
 This is precisely what happens with $\mathscr{U}_{13}$, where the $(1,2)$ and $(2,2)$ modes are added to $\mathscr{U}_9$, showing that the branches with odd parity may also play an important stabilising role.

The methodology is limited in that it will not yield $L$-independent improvements in global stability, but the results in Figure \ref{fig:stability} are stronger than they appear: for $L>4.3$ a different $\tilde{\mathscr{U}}_5$ comprised of the modes from $(0,0)$, $(2,1)$ and $(2,3)$, is found to be globally stable in $A_{(2,1)}$, and the same with the analogous $\tilde{\mathscr{U}}_6$.
Despite not sharing some of the shift-invariant properties of $\Rey_E(L)$ (like $\Rey_E(nL_E)=\Rey_E(L_E)$ for $n\in\mathbb{N}$), this still hints at a parametrized family of modes, which could be studied with more analytical approaches, with the ultimate aim of achieving a result holding at a much broader range of $L$.

Beyond 13 modes we run into computational limitations produced by the large memory footprint and long optimisation solve times involved in the construction of the Lyapunov functionals. 
These computational requirements grow very quickly as a function of $m$, as can be seen from Table \ref{tab:cost}.
Note that for each $L$ several solves are required before finding the largest globally stable $\Rey$ at that $L$.
A single solve for a 15-mode system was tested and took 6 days and over 512GB of memory, which is why constructing a full curve is not currently viable.
Other than the development of specialised and efficient distributed-memory parallelizable optimisation algorithms (note that MOSEK is limited to shared-memory systems), alternative computational reformulations of the methodology are desirable and likely needed for the computation of Lyapunov functionals involving larger number of modes.

\vspace{-3mm}
\begin{table}[h]
	\vspace{-4mm}
	\centering
	\caption{Memory usage and average solution time for each optimisation solve, computed with MOSEK v8.0.0.81 in the same machine, as a function of the number of modes in $\mathscr{U}$.}
	\label{tab:cost}
	\begin{tabular}{lccccccc}
		\hline
		Number of modes in $\mathscr{U}$ & 5 & 6 & 7 & 9 & 11 & 13 & 15 \\
		\hline
		Memory usage (GB) & 0.85 & 0.91 & 1.1 & 2.2 & 8.7 & 39 & 550 \\
		Average solve time (s) & 2.4 & 8.1 & 27 & 470 & $5\,700$ & $54\,000$ & $520\,000$ \\
		\hline
	\end{tabular}
	\vspace{-0mm}
\end{table}
\section{Concluding remarks}
\label{sec:conclusions}

Using the SOS-Lyapunov framework originally proposed by \cite{goulartGlobalStabilityAnalysis2012} and refined by \cite{fuentesGlobalStabilityFluid2022a}, we find 2D plane Poiseuille flow to be globally stable beyond the energy stability limit.
For example, at the critical energy-stable streamwise length, $L_E\approx 2.99$, where $\Rey_E\approx 87.59$, the flow is certified to be globally stable up to $Re \approx 106.8$, representing a $22\%$ improvement.
This is the first enhancement of a lower bound for the global stability limit of this flow since Orr \citeyearpar{orrStabilityInstabilitySteady1907a} originally computed $\Rey_E$ more than a hundred years ago, and shows, for the first time, that this laminar flow is certified to be globally stable even at $\Rey$ where transient energy growth is observed.

To achieve this, finitely many eigenmodes are carefully selected from the energy-rate eigenvalue problem (which is solved numerically using a finite element method).
These small mode sets are capable of capturing sufficient features of the nonlinear dynamics of energy growth and subsequent decay.
Several mode sets resulting in successful probes for quartic Lyapunov functionals at $\Rey>\Rey_E$ are proposed, with the smallest having five modes.
These mode sets could be an important ingredient in being able to prove the existence of Lyapunov functionals using more analytical approaches.
In fact, in the recent work of 
\citet{darrow2026quarticlyapunov},
the same five-mode architecture, which they identified independently, undergirds their analytical construction of quartic Lyapunov functionals for this flow.
It remains to study whether the larger mode sets found in this work may be used for similar analytical constructions, or whether the analytical ans\"atze shed light into further reducing computational costs within the SOS-Lyapunov framework.

The computational methodology is limited to certifying global stability in the range of streamwise periods, $L$, that are tested.
Stronger global stability results in $\Rey$ or for a larger range of $L$ values would require larger mode sets, but computational bottlenecks currently prevent us from achieving that goal.
Similar statements are expected for analogous three-dimensional flows, which are of obvious interest for future research.
Thus, as it relates to their computational cost and memory footprint, reformulations of this methodology or development of specialised optimisation solvers resulting in better scaling properties are extremely desirable and are being actively explored.

\begin{bmhead}[Acknowledgments] 
The authors thank David Goluskin for helpful discussions and suggestions on the original manuscript.
FF acknowledges the partial support of the National Center for Artificial Intelligence CENIA FB210017, Basal ANID based in Chile, and the Fondecyt Grant N.~11261732 from ANID in Chile.
All the authors also gratefully acknowledge the partial support of the Office of Naval Research (ONR) award N629092312098.
\end{bmhead}

\begin{appen}

\renewcommand{\theHsection}{A\arabic{section}}

\vspace{-5.5mm}
\section{SOS formulation} \label{sec:sos_system}
For completeness, we collect the definitions of the operators and constants entering the SOS formulation \eqref{eq:SOScompact}:
\begin{align}
	M_i(\boldsymbol{a},q)
	&=
	\frac{\partial V}{\partial a_i}-2\frac{\partial V}{\partial q^2}a_i,
	\qquad
	\Gtilde(\boldsymbol{a},q)
	=
	\sum\nolimits_{i,j,k=1}^m
	\frac{\partial V}{\partial a_i}\big(\mathsfbi{L}_{ij}a_j+\mathsfbi{N}_{ijk}a_j a_k\big)
	+2\kappa q^2\frac{\partial V}{\partial q^2},
	\label{eq:polySOSformulation}\\
	\mathsfbi{L}_{ij}
	&=
	\big\langle \boldsymbol{u}_i,\tfrac{1}{Re}\nabla^2\boldsymbol{u}_j
	-(\boldsymbol{U}\boldsymbol{\cdot}\nabla)\boldsymbol{u}_j
	-(\boldsymbol{u}_j\boldsymbol{\cdot}\nabla)\boldsymbol{U}\big\rangle,
	\qquad
	\mathsfbi{N}_{ijk}
	=
	-\big\langle \boldsymbol{u}_i,(\boldsymbol{u}_j\boldsymbol{\cdot}\nabla)\boldsymbol{u}_k\big\rangle,
	\label{eq:LNtensors}\\[2pt]
	[\mathsfbi{G}_i]_{jk}
	&=
	\big\langle \tilde{\boldsymbol{h}}_{ij},\tilde{\boldsymbol{h}}_{ik}\big\rangle, 
	\qquad C_i = \max_{(x,y)\in\Omega}
	\rho\big(\tfrac12\big(\nabla\boldsymbol{u}_i+(\nabla\boldsymbol{u}_i)^{\mathrm T}\big)\big)
	,
	\label{eq:boundsSOSformulation}\\
	\boldsymbol{h}_{i0}
	&=
	\tfrac{1}{Re}\Delta\boldsymbol{u}_i
	+(\boldsymbol{U}\boldsymbol{\cdot}\nabla)\boldsymbol{u}_i
	-(\nabla\boldsymbol{U})^{\mathrm T}\boldsymbol{u}_i,
	\qquad
	\boldsymbol{h}_{ij}
	=
	(\boldsymbol{u}_j\boldsymbol{\cdot}\nabla)\boldsymbol{u}_i
	-(\nabla\boldsymbol{u}_j)^{\mathrm T}\boldsymbol{u}_i,
	\label{eq:auxiliaryGbounds}
\end{align}
where $\kappa$ is the largest energy eigenvalue not included in $\mathscr U$, $\rho(\cdot)$ denotes the spectral radius, and $\tilde{\boldsymbol{h}}_{ij}$ are the solenoidal projections of $\boldsymbol{h}_{ij}$ onto the orthogonal complement of $\mathscr U$. See \cite{fuentesGlobalStabilityFluid2022a} for derivation.
\vspace{-3mm}
\section{Gram matrix calculation}\label{app:solenoidal_projection}

To construct the Gram matrices $\mathsfbi G_i\in\mathbb{R}^{(m+1)\times(m+1)}$,
first note that each $\boldsymbol h_{ij}$ in \eqref{eq:auxiliaryGbounds} has a Helmholtz decomposition of the form ${\boldsymbol h}_{ij}={\boldsymbol h}_{ij}^{\mathrm{div}}+\nabla \phi$, where ${\boldsymbol h}_{ij}^{\mathrm{div}}$ is an $x$-periodic divergence-free solenoidal projection (i.e. $\nabla\boldsymbol{\cdot} {\boldsymbol h}_{ij}^{\mathrm{div}} = 0$) of $\boldsymbol{h}_{ij}$, in this case uniquely determined by the boundary conditions ${\boldsymbol h}_{ij}^{\mathrm{div}}\boldsymbol{\cdot} \boldsymbol n=0 $ at $y=\pm 1$.
To compute it, note that the $x$-periodic scalar potential $\phi$ satisfies the following Poisson equation,
\begin{equation}
\nabla^2 \phi = \nabla\boldsymbol{\cdot} {\boldsymbol h}_{ij}
\qquad\text{with}\qquad
\nabla \phi\boldsymbol{\cdot} \boldsymbol n = {\boldsymbol h}_{ij}\boldsymbol{\cdot} \boldsymbol n
\quad\text{on } y=\pm 1.
\label{eq:poissonproblem}
\end{equation}
Letting $\phi(x,y) = \sum_{n\in\mathbb Z} \varphi_n(y) e^{\text{i}\alpha_n x}$ with $\alpha_n = \frac{2\pi n}{L}$ allows \eqref{eq:poissonproblem} to be projected for each $n$ and recast as $\varphi_n''-\alpha_n^2\varphi_n=f_{ij,n}$ with Neumann boundary conditions $\varphi_n'(\pm1)=g_{ij,n}^{\pm}$.
Here, 
 \begin{equation}
f_{ij,n}(y)
=
\textstyle\frac{1}{L}\int_0^L
\nabla\!\boldsymbol{\cdot}\!{\boldsymbol h}_{ij}(x,y)\,e^{-\text{i}\alpha_n x}\,\mathrm dx
\quad\text{and}\quad
g_{ij,n}^{\pm}
=
\frac{1}{L}\int_0^L
({\boldsymbol h}_{ij})_2(x,\pm1)\,e^{-\text{i}\alpha_n x}\,\mathrm dx.
\end{equation}
When $n\neq 0$, this equation has a unique solution, but when $n=0$ the Neumann problem determines $\varphi_0$ only up to an additive constant, so we fix this by imposing $\varphi_0(-1) = 0$ (this choice is immaterial, since only $\nabla \phi$ enters the construction). 
In either case we choose to multiply by a test function and integrate by parts to produce variational formulations of the form, 
\begin{equation}
\begin{gathered}
    \text{find}\,\,\varphi\in \mathscr{W}\,\,\text{such that}\quad  
    \mathscr a_n^{\boldsymbol h}(\varphi, \eta) = \ell_n^{\boldsymbol h}(\eta)\quad \forall \eta\in \mathscr{W}, \quad\text{where}\\
     \mathscr a_n^{\boldsymbol h}(\varphi, \eta) = \textstyle\int_{-1}^1
(
\varphi'\overline{\eta}'
+\alpha^2_n\varphi\overline{\eta}
)\,\mathrm dy,\quad\ell_n^{\boldsymbol h}(\eta) = -\int_{-1}^1 f_{ij,n}\overline{\eta}\,\mathrm dy
+g_{ij,n}^{+}\overline{\eta}(1)
-(1-\delta_{0n})g_{ij,n}^{-}\overline{\eta}(-1)\,.
\end{gathered}
\label{eq:weakformssolenoidal}
\end{equation}
When $n\neq0$, $\mathscr{W}$ is selected as the Sobolev space $H^1(-1, 1)$, whereas when $n=0$ we select $\mathscr{W}$ as the subset of $H^1(-1,1)$ whose functions vanish at $y=-1$.
Note \eqref{eq:weakformssolenoidal} must only be solved for the values of $n$ present in $\boldsymbol{h}_{ij}$ (which typically involves only very few modes), after which we reconstruct $\phi$ and ${\boldsymbol h}_{ij}^{\mathrm{div}} = \boldsymbol{h}_{ij} - \nabla\phi$.

Lastly, we project ${\boldsymbol h}_{ij}^{\mathrm{div}}$ to the space orthogonal to $\operatorname{span}(\mathscr U)$, i.e., 
\begin{equation}
\tilde{\boldsymbol h}_{ij}
=
{\boldsymbol h}_{ij}^{\mathrm{div}}
-
\textstyle\sum_{k=1}^m
\langle {\boldsymbol h}_{ij}^{\mathrm{div}},\boldsymbol u_k\rangle
\,\boldsymbol u_k,
\label{eq:orthonormalproyection}
\end{equation}
since $\mathscr U = \{\boldsymbol{u}_1,\ldots,\boldsymbol{u}_m \}$ is a solenoidal orthonormal set, and then compute $[\mathsfbi G_i]_{jk}=\langle \tilde{\boldsymbol h}_{ij},\tilde{\boldsymbol h}_{ik}\rangle$.

To numerically solve \eqref{eq:weakformssolenoidal} we use a finite element method by discretizing $\mathscr W$ using Hermite finite elements as described in \S\ref{sec:nummethod}, while the remaining integrals in \eqref{eq:orthonormalproyection} and $\mathsfbi{G}_i$ are exactly computed numerically. 
\vspace{-3mm}
\section{Spectral radii calculation}\label{app:spectral_rad}

To compute the $C_i$ associated with $\boldsymbol{u}_i=(\partial_y \psi_i,-\partial_x\psi_i)$ and specified in \eqref{eq:boundsSOSformulation}, we use that $\nabla\boldsymbol{u}_i$ has zero trace (due to its incompressibility) to derive,
\begin{equation}
    C_i= \textstyle{\max_{(x,y)\in\Omega}}
    \,\rho\big(\tfrac{1}{2}\big(\nabla\boldsymbol{u}_i+
        \left(\nabla\boldsymbol{u}_i\right)^{\mathrm T}\big)\big)
        =\textstyle{\max_{(x,y)\in\Omega}}
            \sqrt{(\partial_{xy}^2\psi_i)^2
            +\tfrac{1}{4}(\partial_y^2\psi_i-\partial_x^2\psi_i)^2}.
\end{equation}
(Note there is a typo in (S16) of \cite{fuentesGlobalStabilityFluid2022a}, but the formula above, which underlies the actual computations in \cite{fuentesGlobalStabilityFluid2022a} and in this work, is the correct one.)
To avoid a 2D optimisation over $\Omega$, the idea is to fix $y$ and analytically optimize over $x$ first, exploiting the periodicity in this direction. 
Assuming the $\boldsymbol{u}_i$ are energy eigenmodes, then $\psi_i(x,y) = \mathfrak{R}(\varphi(y)e^{\text{i}\alpha x})$ for some $\varphi$ and $\alpha$.
We leave as an exercise to the reader to show that
\begin{gather}
    (\partial_{xy}^2\psi_i)^2
            +\tfrac{1}{4}(\partial_y^2\psi_i - \partial_x^2\psi_i)^2=A(y) + B(y)\cos(2\alpha x - \delta(y))\qquad\text{with}\quad 
            \label{eq:Cisqexact}\\
    A(y)=\tfrac{1}{2}\big(\beta(y)^2+\gamma(y)^2\big),\qquad
    B(y)=\tfrac{1}{2}\sqrt{\beta(y)^4+\gamma(y)^4
        +2\beta(y)^2\gamma(y)^2\cos\big(2(\theta_1(y)-\theta_2(y))\big)}, \\
    \beta(y)=\alpha |\varphi'(y)|,\quad 
        \gamma(y)=\tfrac{1}{2}|\varphi''(y)+\alpha^2\varphi(y)|,\\
        \theta_1(y) = \arg\!\big(\text{i}\alpha\varphi'(y)\big), 
        \quad \theta_2(y) 
            = \arg\!\big(\varphi''(y)+\alpha^2\varphi(y)\big),
\end{gather}
where $\delta(y)$ is some angle. 
The maximum of \eqref{eq:Cisqexact} in $x\in(0,L)$ is attained when $\cos(2\alpha x - \delta(y))=1$, so that
\begin{equation}
    C_i^2=\textstyle{\max_{-1\leq y\leq1}}A(y)+B(y)\,.
\end{equation}
Consequently, we have reduced this to a one-dimensional optimisation problem that is much easier to solve.
\vspace{-3mm}
\section{Verification and other implementation details}
\label{app:implementation} 

To reduce the computational cost, we restrict the ans\"atze of $P$, $r_i$ and $s_i$ in \eqref{eq:SOScompact} by forcing them to respect the symmetries of the truncated dynamics governed by $\tfrac{\mathrm d a_i}{\mathrm d t}=\sum_{j=1}^m\mathsfbi{L}_{ij}a_j+\sum_{j,k=1}^m\mathsfbi{N}_{ijk}a_j a_k=\boldsymbol{f}_i(\boldsymbol{a})$ (for $\mathsfbi{L}$ and $\mathsfbi{N}$ in \eqref{eq:LNtensors}).
This imposes linear relations between the coefficients of the monomials in the ans\"atze (like sign symmetries), which are precisely the optimisation parameters, often resulting in some coefficients outright vanishing and in a significant reduction in the number of free parameters. 
The symmetries themselves are dictated by the flow and choice of mode set $\mathscr U$, but ultimately, the idea is that for each equivariant symmetry $\Psi$ of the dynamical system (meaning a linear isometry $\Psi:\mathbb{R}^m\to\mathbb{R}^m$ with $\Psi^K=\mathsfbi{I}$ for some $K\in\mathbb{N}$ satisfying $\Psi(\boldsymbol{f}(\boldsymbol{a}))=\boldsymbol{f}(\Psi(\boldsymbol{a}))$), we enforce that $P(\Psi(\boldsymbol{a}),q)=P(\boldsymbol{a},q)$ and the same with the $r_i$ and $s_i$.
For this flow, if $(a_j,a_{j+1})$ are the pair of coefficients of the eigenmodes associated with $(n,k)$ for $n\neq0$ (see Figure \ref{fig:EnergyEVP}), the dynamics are invariant under $\mathcal{R}^n$ where $\mathcal{R}(a_j)=a_{j+1}$, $\mathcal{R}(a_{j+1})=-a_j$, and $\mathcal{R}(a_k)=a_k$ for $k\notin\{j,j+1\}$ (note that $\mathcal{R}^4=\mathsfbi{I}$).

For example, for the 5-mode set $\mathscr{U}_5$ described in \S\ref{sec:results}, one such equivariant symmetry (of `order' $K=4$) is described by the isometry $(a_1, a_2, a_3, a_4, a_5) \mapsto (a_1, -a_3, a_2, -a_5, a_4)$. 
Notice that the monomial $a_2a_3$ is mapped to $-a_2a_3$ under this transformation, so if $P(\Psi(\boldsymbol{a}),q)=P(\boldsymbol{a},q)$, then clearly the monomial $a_2a_3$ cannot be present in $P$ (i.e., its monomial coefficient is zero).
Similarly, $a_3a_5$ is mapped to $a_2a_4$, and $a_2a_4$ is mapped back to $a_3a_5$, so the invariance implies both monomials must have the same coefficient, resulting in an invariant term of the form $a_3a_5+a_2a_4$ in $P$ accompanied by the same coefficient. 
Likewise, the combination $a_3a_4-a_2a_5$ is invariant, as is $\boldsymbol{a}\boldsymbol{\cdot}\boldsymbol{a}$. 
As an illustration, at $L=2.99$ and $\Rey=92.3$, a feasible SDP solution results in the following Lyapunov functional,
\begin{equation}
	V(\boldsymbol a, q)= E^2 + 0.215 a_1 E+ 0.0283 E -0.00131 (a_3a_5+a_2a_4) - 0.000271 (a_3a_4-a_2a_5),
	\label{eq:example}
\end{equation}
where $E=\tfrac{1}{2}(\boldsymbol{a}\boldsymbol{\cdot}\boldsymbol{a}+q^2)$ and only three significant digits of each coefficient are being displayed.
In this case, the Lyapunov functional shares the same structure as the one proposed by \cite{darrow2026quarticlyapunov} for $\mathscr{U}_5$, but, in principle, our ansatz can vary more freely and is applicable to any $\mathscr{U}_m$ (which, in practice, may come at the cost of obfuscating a possible physical interpretation).

\begin{figure}
	\centering
	\includegraphics[width=\linewidth]{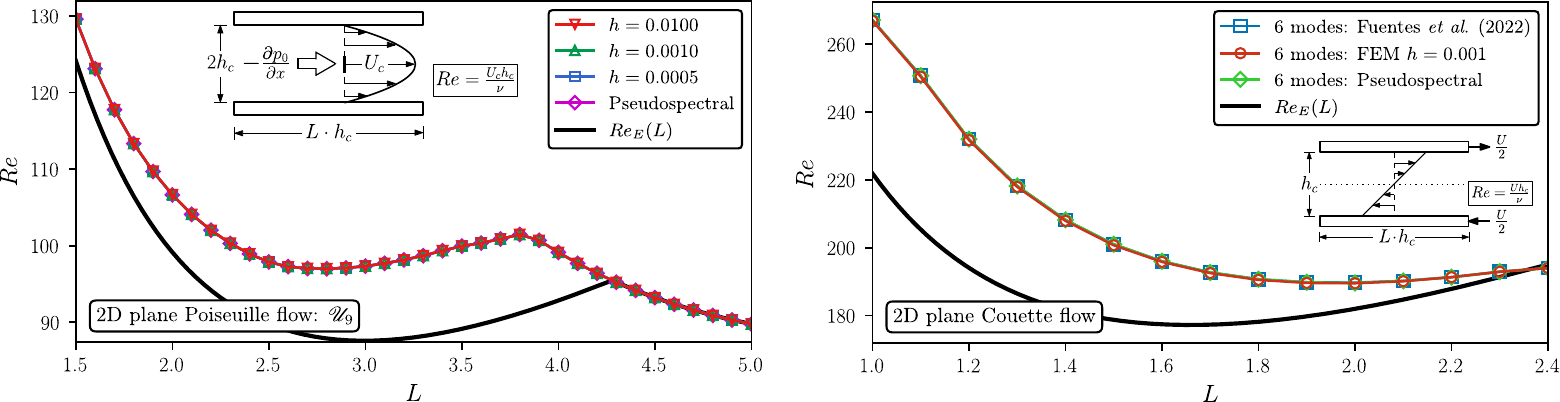}
	\caption{
		The left panel shows the nearly indistinguishable feasibility thresholds using the mode set $\mathscr{U}_9$ (see Figure \ref{fig:EnergyEVP} for definition) for the FEM implementation with different mesh sizes as well as the implementation using pseudospectral methods.
		The right panel shows that the results of the FEM and pseudospectral implementations of 2D plane Couette flow match the data from \cite{fuentesGlobalStabilityFluid2022a}, which are taken as the benchmark.
	}
	\label{fig:verification}
	\vspace{-3mm}
\end{figure}

The energy eigenvalue problem \eqref{eq:energy2D} (in fact, both \eqref{eq:energyEVP} and \eqref{eq:energyEVP0}), along with the auxiliary Poisson equations in Appendix \ref{app:solenoidal_projection}, are solved using FEM with element size $h=0.001$, as noted in the main text.
To verify this mesh size was sufficiently small we performed a mesh convergence analysis on specific extreme test cases, complemented by visual inspection to ensure all solution details were resolved.
Moreover, we solved the entire global stability problem with $\mathscr{U}_9$ (see Figure \ref{fig:EnergyEVP} for its definition) for $h=0.01$, $h=0.001$ and $h=0.0005$, achieving essentially the same results in all three cases with a maximum discrepancy in $\Rey$ of 0.025, as can be appreciated in Figure \ref{fig:verification}.
In truth, $h=0.01$ is probably sufficiently accurate, and an adaptive mesh with less elements could also attain the same level of accuracy, but for simplicity we use uniform $h=0.001$ throughout.
Lastly, we used the results of 2D plane Couette flow from \cite{fuentesGlobalStabilityFluid2022a} with a 6-mode $\mathscr{U}$ as a benchmark to compare against the independently-coded global stability computations via FEM described in this work.
The results are also nearly indistinguishable, as observed in Figure \ref{fig:verification}.

As a completely separate endeavour, we coded and solved \eqref{eq:energy2D} and the bounds in Appendix \ref{app:solenoidal_projection} using ultraspherical pseudospectral methods.
The results were then verified against those of FEM in both 2D plane Couette and plane Poiseuille flows, also getting nearly indiscernible results in the aforementioned test cases (see Figure \ref{fig:verification}).
That said, sometimes, for a fixed $L$, we occasionally observed small isolated intervals of apparent infeasibility bracketed by feasible Reynolds numbers, which we interpreted as false negatives. 
These events were nearly non-existent when using the FEM implementation compared to the pseudospectral one.
Thus, all results reported in the main text involve FEM, since it yielded much more robust and consistent results.  

Near the stability boundary, in rare cases, repeated runs of the SDP solver occasionally returned different feasibility outcomes for the exact same inputs.
We mitigated this sensitivity by checking neighbouring Reynolds numbers. 
This behaviour may be related to near-degeneracies in the SDP and the possible use of random initial guesses in the optimisation algorithms.
The use of a fixed streamwise shift convention for the $n\neq0$ energy eigenmodes (see \S\ref{sec:modes}, where we ensure a stagnation point at the origin for one of the eigenmodes) may provide some numerical stability, because, when this was left unspecified, we observed minor discrepancies, typically below $\pm0.1\Rey$.
Lastly, the use of an `old' version of MOSEK, namely v8.0.0.81, is due to a tuning change in v8.1 and above, which results in a memory footprint that is simply too large for modern machines.

Feasibility (or infeasibility) of the SDP was determined using the stopping criteria of the conic interior-point solver in MOSEK, with a solution being classified as feasible only when MOSEK returned the status \texttt{PRIMAL\_AND\_DUAL\_FEASIBLE}.
We refer to \S9 of the manual for the MOSEK v8.0 toolbox for MATLAB \citeyearpar{mosek} for more details (or to \S13.3 for the version v11.2.2).
The default termination (relative) tolerances would likely work well, but we used slightly stricter tolerances given by $\varepsilon_p = 10^{-11}$ for primal feasibility, $\varepsilon_d = 10^{-11}$ for dual feasibility, $\varepsilon_g = 10^{-11}$ for the relative gap, $\varepsilon_i = 10^{-15}$ for infeasibility, $10^{-11}$ for the complementarity gap, and $10^5$ for the parameter \texttt{MSK\_DPAR\_INTPNT\_CO\_TOL\_NEAR\_REL}.

\end{appen}

\bibliographystyle{jfm}
\bibliography{refs}

\end{document}